\documentclass[12pt]{article}
\usepackage{amssymb, amsmath, xspace, lscape,  latexsym,   
}
\renewcommand{\dim}{\mathrm{dim \,}}

\renewcommand{\le}{\leqslant}

\renewcommand{\ge}{\geqslant}

\newcommand{\vf}{\varphi}

\newcommand{\lb}[1]{\label{#1}}

\newcommand{\mref}[1]{(\ref{#1})}

\newcommand{\p}{\partial}

\renewcommand{\a}{\alpha}

\renewcommand{\b}{\beta}

\newcommand{\g}{\gamma}

\newcommand{\C}{\mathbb C}

\newcommand{\N}{\mathbb N}

\newcommand{\anm}{{{\cal A}_n(m)}}

\newcommand{\adva}{{{\cal A}_2(m)}}

\newcommand{\cdva}{{{\cal C}_{2}(m,l)}}

\newcommand{\cnm}{{{\cal C}_{n+1}(m,l)}}

\renewcommand{\hat}{\widehat}

\renewcommand{\tilde}{\widetilde}

\def\beq#1#2\eeq{%
            \begin{equation}%
            \label{#1}%
                #2%
            \end{equation}%
        }

\def\btheor#1\etheor{%

         \begin{theor}%

             #1%

         \end{theor}%
     }

     \def\bsled#1\esled{%
         \begin{sled}%
             #1%
         \end{sled}%
     }

\newcommand{\cl}{{\cal L}}

\newcommand{\cA}{{\cal A}}

\newtheorem{theorem}{Theorem}

\newtheorem{lemma}{Lemma}
\newtheorem{prop}{Proposition}
\newtheorem{cor}{Corollary}

\newcommand{\nad}[2]{\genfrac{}{}{0pt}{}{#1}{#2}}

\def\hm#1{#1\nobreak\discretionary{}{\hbox{\m@th$#1$}}{}}
\def\mi#1{\discretionary{\hbox{\m@th$#1$}}{\hbox{\m@th$#1$}}{}}

\begin{document}

\begin{center}

{\bf\Large Quasi-invariants and quantum integrals of
the deformed Calogero--Moser systems}

\vspace{1cm}

{\bf  M. Feigin $^1$ and A. P. Veselov $^{2,3}$}
\end{center}

\vspace{1cm}

\noindent $^1$ Chair of Mathematics and Financial Applications,
Financial Academy, \\ Leningradsky prospect, 49, Moscow, 125468, Russia

\noindent $^2$ Department of Mathematical Sciences, Loughborough
University,\\ Loughborough,  LE11 3TU, UK

\noindent $^3$ Landau Institute for Theoretical Physics, Kosygina
2, Moscow, 117940, Russia

\noindent E-mail addresses: mfeigin@dnttm.ru,
A.P.Veselov@lboro.ac.uk

\vspace{1cm}

\begin{abstract}
\noindent The rings of quantum integrals of the generalized
Calogero-Moser systems related to the deformed root systems ${\cal A}_n(m)$
and ${\cal C}_n(m,l)$ with integer multiplicities and corresponding algebras
of quasi-invariants are investigated. In particular, it is shown
that these algebras are finitely generated and free as the modules
over certain polynomial subalgebras (Cohen-Macaulay property). The
proof follows the scheme proposed by Etingof and Ginzburg in the
Coxeter case. For two-dimensional systems the corresponding
Poincare series and the deformed $m$-harmonic polynomials are
explicitly computed.
\end{abstract}


\section{Introduction}
Quantum Calogero-Moser(CM) problem \cite{C3} with the Hamiltonian
\beq{Cal} L = \Delta- \sum_{i<j}^n \frac{2m(m+1)}{(x_i-x_j)^2}
\eeq was generalized by Olshanetsky and Perelomov \cite{OP1} for
any Coxeter group as \beq{OPdef} L=\Delta-\sum_{\a\in
{\cA}}\frac{m_\a(m_\a+1)(\a,\a)}{(\a,x)^2} \eeq where $\cA =
\cal{R}_+$ is a positive part of the corresponding root system. In
1996 O. Chalykh and the authors \cite{CFV1} showed that there are
non-Coxeter integrable generalizations as well (so-called deformed
Calogero-Moser problems). Recently A.N. Sergeev suggested an
explanation of these deformations in relation with Lie
superalgebras \cite{Ser}. A systematic approach to the deformed
quantum CM problems from this point of view has been developed in
\cite{SV}.

In this paper, which can be considered as a sequel to
\cite{FV1,FV2}, we investigate the algebras of quantum integrals
for two series $\anm$ and $\cnm$ of the deformed quantum
Calogero-Moser systems discovered in \cite{CFV2,CFV3}. According
to \cite{SV} these series are the only non-Coxeter cases among the
deformed CM systems when all the parameters are integer.

The corresponding operators have the forms \cite{CFV2,CFV3}
\beq{AApot} L=\Delta-\sum_{i<j}^n
\frac{2m(m+1)}{(x_i-x_j)^2}-\sum_{i=1}^n\frac{2(m+1)}{(x_i-\sqrt m
x_{n+1})^2}, \eeq and \label{Cpot}
\begin{eqnarray}
L & = & \Delta -  \sum_{i<j}^n \frac{4\kappa(\kappa+1)(x_i^2 +
x_j^2)}{(x_i^2 - x_j^2)^2} - \sum_{i=1}^n \frac{m(m+1)}{x_i^2
}-\nonumber\\ &&\label{CC1}\\ && \frac{l(l+1)}{x_{n+1}^2} -
\sum_{i=1}^n \frac{4(\kappa+1)(x_i^2 + \kappa
  x_{n+1}^2)}{(x_i^2 - \kappa x_{n+1}^2)^2},\nonumber
\end{eqnarray}
where the parameters $m, \kappa, l$ satisfy the relation
$\kappa=\frac{2m+1}{2l+1}$ and $\Delta$ stands for the standard
Laplace operator in $n+1$-dimensional Euclidean space. When the
parameter $m=1$ the first operator becomes a special case of the
Calogero operator (\ref{Cal}). The second operator is the
deformation of the generalized CM operator (\ref{OPdef}) related
to the root system ${\cal C}_{n+1}$, which corresponds to the case
$m=l$.

In this paper we consider the case when all the parameters
(multiplicities) in these operators are integer. The importance of
the last condition was first demonstrated by Chalykh and one of
the authors \cite{CV}, who discovered that in such a case the ring
of integrals of quantum CM system is much bigger than for the
generic parameters.

More precisely, for any configuration $\cA$ which is a finite set
of vectors $\a$ in the Euclidean space $V$ with prescribed
multiplicities $m_{\alpha} \in {\bf Z}_+$ one can introduce the
following algebra of {\it quasi-invariants} $Q = Q^\cA$. It
consists of all polynomials $q$ on $V$ with the following
property: \beq{quasidefIntr} q(s_\a(x))=q(x) +
o\left((\a,x)^{2m_\a}\right) \eeq near the hyperplane $(\a,x)=0$
for any $\a\in \cA$, where $s_\a$ denotes reflection with respect
to this hyperplane. For the special configurations, in particular
for the Coxeter systems and configurations $\anm$ and $\cnm$, there exists a homomorphism from the
algebra of quasi-invariants into the ring of quantum integrals of
the corresponding generalized CM operator, which under some
assumptions can be shown to be an isomorphism.

For the Coxeter configurations the algebraic structure of the
rings of quasi-invariants $Q^\cA$ was investigated in
\cite{FV1,FV2,EG}. In particular, Etingof and Ginzburg \cite{EG}
proved that this ring is a free module over the subring of
invariant polynomials (Cohen-Macaulay property) confirming some of
the conjectures from \cite{FV1}.

The main result of this paper is the proof of a similar fact for
the rings of quasi-invariants related to the deformed
configurations $\anm$ and $\cnm.$ We introduce certain polynomial
subalgebra $P^\cA \subset Q^\cA$ and show that $Q^\cA$ is free as
a module over $P^\cA$. The proof follows the Etingof-Ginzburg
scheme from \cite{EG}. In two-dimensional case we find the
explicit formulas for the Poincare series of the corresponding
rings of quasi-invariants and show that these rings are
Gorenstein. We introduce also the deformed $m$-harmonic
polynomials and compute them explicitly in ${\cal A}_2(m)$-case.

\section{Deformed quantum Calogero-Moser systems}


The quantum systems we are going to discuss are related to the
following configurations introduced in \cite{CFV1,CFV2,CFV3}.

The first configuration $\anm$ consists of the vectors $e_i-e_j$
with multiplicity $m$, where $1\le i<j \le n$, and the vectors
$e_i-\sqrt m e_{n+1}$ with multiplicity 1. When the parameter
$m=1$ this is the classical root system of type $\cA_n$.

The second configuration $\cA=\cnm$ consists of the following
vectors
$$
\label{C}
\cnm =
\left\{
\begin{array}{ll}
e_i\pm e_j &  {\rm with \,\, multiplicity \,\,}   \kappa\\ 2e_i &
{\rm with \,\, multiplicity \,\,}   m\\ 2\sqrt{\kappa}e_{n+1} &
{\rm with \,\, multiplicity \,\,}  l\\ e_i\pm \sqrt{\kappa}e_{n+1}
& {\rm with \,\, multiplicity \,\,}  1\\
\end{array}
\right.
\nonumber
$$
where $m, \kappa$ and $l$ are parameters with the relation
$\kappa=\frac{2m+1}{2l+1}$ (so only two of them, say $m,l$ are
independent) and $1\le i<j\le n$. In the case of ${\cal C}_2
(m,l)$ system there are no vectors of $e_i\pm e_j$ type, and the
parameters $m,l$ can be arbitrary. In the case $l=m$ the system
$\cnm$ coincides with the classical root system ${\cal C}_{n+1}$
(or ${\cal D}_{n+1}$ for $l=m=0$).

Although in the rest of the paper only the case of integer
multiplicities will be considered at the beginning we will not
assume this and consider general values of parameters.
Corresponding {\it deformed quantum CM problems} are given by the
general formula (\ref{OPdef}) or more explicitly by the formulas
(\ref{AApot}), (\ref{CC1}) respectively.

We will actually be using these operators in a different
("radial") gauge and consider the operators $\cl = g L g^{-1}$
with $g=\prod_{\a\in \cA}(\a,x)^{m_\a}$, which have the following
forms:

\beq{AA}
{\cal L}_{\cA_n(m)} = \Delta
-\sum_{i<j}^n\frac{2m}{x_i-x_j}(\p_i-\p_j)-
\sum_{i=1}^n\frac{2}{x_i-\sqrt m x_{n+1}}(\p_i-\sqrt m\p_{n+1}).
\eeq
and
\begin{eqnarray}
\label{CC11} {\cal L}_{C_{n+1}(m,l)}  & = & \Delta - \sum_{i<j}^n
\frac{4\kappa(x_i\p_i - x_j\p_j)}{x_i^2 - x_j^2} - \sum_{i=1}^n
\frac{2m\p_i}{x_i}-\nonumber\\ &&\label{CC}\\ &&
\frac{2l\p_{n+1}}{x_{n+1}} - \sum_{i=1}^n \frac{4(x_i\p_i - \kappa
  x_{n+1}\p_{n+1})}{x_i^2 - \kappa x_{n+1}^2}.\nonumber
  \end{eqnarray}
The existence of such forms for $\anm$ and $\cnm$ is one of the
remarkable properties of these systems and is due to the following
identity valid for both systems: \beq{firstid}
\sum_{\nad{\a\ne\b}{\a,\b\in \cA}} \frac{m_\a m_\b
(\a,\b)}{(\a,x)(\b,x)} \equiv 0, \eeq which is equivalent to the
set of relations \beq{firstidlist} \sum_{\nad{\b\ne\a}{\b\in \cA}}
\frac{m_\b (\a,\b)}{(\b,x)} \equiv 0 \quad \text{at} \quad
(\a,x)=0 \eeq for any $\a\in \cA$.

As it was shown in \cite{CFV1,CFV3} the quantum systems related to
the deformations $\anm$ and $\cnm$ are integrable. More precisely,
consider the following two series of polynomials \beq{ps}
p_s=x_1^s+ x_2^s + \ldots + x_n^s + m^{\frac{s-2}2} x_{n+1}^s,
\eeq and \beq{qs} q_s=x_1^{2s}+ x_2^{2s} + \ldots + x_n^{2s} +
\kappa^{s-1} x_{n+1}^{2s}, \eeq where $s=1,2,\ldots$
\begin{theorem}\cite{CFV3}\lb{teor1}
For any $s\in \N$ there exists differential operator $\cl_s$ with
the highest term $p_s(\p)$ given by \mref{ps} such that
$[\cl_s,\cl_t]=0$, $t\in \N$. The operator $\cl_2$ coincides with
the Calogero--Moser operator \mref{AA} related to the system
$\anm$. The same is true for the polynomials $q_s$ and the
operator \mref{CC} related to $\cnm.$
\end{theorem}
Thus for generic parameters we have a commutative algebra of
quantum integrals of the deformed CM problems generated by
$\cl_s,$ which is isomorphic to the subalgebra generated by the
polynomials $p_s$ in $\anm$ case and by $q_s$ in $\cnm$ case. One
can show that these subalgebras are finitely generated for generic
values of the parameters (see \cite{SV}).

However we will be using the smaller subalgebras $P^{\anm} = {\bf
C}[p_1, p_2, \dots, p_{n+1}] $ and $P^{\cnm} =  {\bf C}[q_1, q_2,
\dots, q_{n+1}],$ which are freely generated by the first $n+1$
polynomials $p_s$ and $q_s$ respectively. Since when $m=1$ and
$\kappa=1$ these algebras coincide with the corresponding algebras
of invariants for the Coxeter groups of type $A_{n+1}$ and
$C_{n+1}$ we will call them the {\it algebras of deformed
invariants.}

\begin{prop}\label{prop1}
If parameter $m\notin\{0, -1,
-\frac12,-\frac13,\ldots,-\frac1{n}\}$ then
\begin{enumerate}
\item
The dimension of the quotient $$\C[x_1,\ldots,x_{n+1}]/I,$$ where
$I$ is ideal generated by $p_1,\ldots,p_{n+1},$ is finite and
equals to the number $\mu^\anm =(n+1)!$ of different complex
solutions to the algebraic system $p_i(x)=c_i$, $i=1,\ldots, n+1$
for generic $c_i \in {\bf C}.$
\item
The ring $\C[x_1,\ldots,x_{n+1}]$ is a free module over
$\C[p_1,\ldots,p_{n+1}]$ of rank $\mu^\anm$.
\end{enumerate}
The same is true for the system $\cnm$ if we replace $m$ by
$\kappa$, $p_s$ by $q_s$ and $\mu^\anm$ by $\mu^\cnm =
2^{n+1}(n+1)!$.
\end{prop}
For the proof we need the following
\begin{lemma}\lb{isol0}
The system of equations
$$
p_s(x)=0, \quad s=1, \ldots, n+1
$$
  has
unique solution $x=0$ if $p_s$ are given by \mref{ps} where
$m\notin\{0, -1, -\frac12,-\frac13,\ldots,-\frac1{n}\}$.
\end{lemma}
{\bf Proof} Let us scale the variable $x_{n+1}$ introducing
$y=m^{1/2}x_{n+1}$. Then the system of equations takes the form
\begin{equation}\label{sJack}
\begin{cases}
x_1+\ldots+x_n+m' y=0\\
x_1^2+\ldots+x_n^2+m' y^2=0\\
\hbox to 15mm {\dotfill} \relax\\
x_1^{n+1}+\ldots+x_n^{n+1}+m' y^{n+1}=0
\end{cases}
\end{equation}
where $m'=\frac1{m}$.
We have to show that \mref{sJack} implies that $x_1=\ldots
=x_n=y=0$. One can see that this is not the case when
$m'=0,-1,-2,\ldots, -n$. Indeed, if $m'=0$ then there exists
nonzero solution $x_i=0$, $i=1,\ldots,n$, and $y\in\C$ is
arbitrary. If $m'=-s$ then the nonzero solution will be
$x_1=x_2=\ldots=x_s=y\in\C$ and $x_{s+1}=\ldots = x_n=0$. We will
show that there are no nonzero solutions for all other values of
$m'$.

Denote by $\tau_i$ the power sum $$ \tau_i=x_1^i+\ldots+x_n^i. $$
Then $\tau_1, \ldots, \tau_n$ form a basis in the ring of
symmetric functions of $n$ variables $x_1,\ldots,x_n,$ in
particular, $\tau_{n+1}=P(\tau_1,\ldots,\tau_n)$ for some
polynomial $P(a_1,\ldots,a_n)$. If we assign degree $i$ to the
variable $a_i$ then $P$ is homogeneous polynomial of degree $n+1$.
The first $n$ equations of \mref{sJack} can be written as
$\tau_i=-m'y^i$. Then the last equation of the system can be
written as $$ P(-m'y, -m'y^2,\ldots,-m'y^{n})+m' y^{n+1}=0. $$ Now
due to weighted homogeneity of $P$, $$ P(-m'y,
-m'y^2,\ldots,-m'y^{n})=y^{n+1}P(-m', -m',\ldots,-m'), $$ and
usual degree $\deg P\le n+1$. It is obvious from \mref{sJack} that
if nonzero solution exists then $y\ne 0$. Therefore $m'$ must
satisfy $$ P(-m', -m',\ldots,-m') +m'=0 $$ with $\deg P\le n+1$.
Hence either $m'$ is arbitrary or there are not more than $n+1$
possible values for $m'$ to have nonzero solution to \mref{sJack}.
Since for $m'=1$ there are no nonzero solutions, and since we know
that for $m'=0,-1,\ldots,-n$ there are nonzero solutions, we
conclude that if $m'\ne 0, -1, \ldots, -n$ then there are no
nonzero solutions to system \mref{sJack}. Lemma is proven.

Now the part 1 of Proposition 1 for the system $\anm$ follows from
the Lemma and the standard results about isolated zeros of
analytic maps \cite{AVG}. The part 2 is a consequence of part 1
and the Cohen--Macaulay property of the polynomial ring. For
$\cnm$-system the proof is similar.

Let now ${\cA}$ denote one of the systems $\anm$ or $\cnm$ with
the parameters $m$ and $\kappa$ which do not belong to the
exceptional set $\{0, -1, -\frac12,-\frac13,\ldots,-\frac1{n}\}$
and let $\g_1(x)=1,\g_2(x),\ldots,\g_{\mu^{\cA}}(x)$ be some
homogeneous basis in $\C[x_1,\ldots,x_{n+1}]$ as a module over the
corresponding algebra of the deformed invariants $P^{\cA}.$

Consider the system of the differential equations \beq{sys} {\cal
L}_i f = \lambda_i f, \quad i=1,\ldots, n+1, \eeq where $\cl_i =
\cl_i^\cA$ are the commuting quantum integrals defined in Theorem
\ref{teor1}. Let $F(x)$ be the vector function $$
F(x)=(f(x),\g_2(\p)f(x),\ldots,\g_{\mu^{\cA}}(\p)f(x)).$$
\begin{prop}\label{prophol}
The system of equations \mref{sys} is equivalent to the first
order system $$ \frac{\p F}{\p x_i} = A_i(x,\lambda) F, \qquad
i=1,\ldots,n+1, $$ where $A_i(x)$ are matrix valued functions
analytic if $\prod_{\a\in {\cA}} (\a,x)\ne 0$ and such that $$
[\p_i-A_i,\p_j-A_j]=0. $$ The space of local analytic solutions to
\mref{sys} has the dimension $\mu^{\cA}$.
\end{prop}

For the root systems such a statement (in a trigonometric version)
was proven in \cite{HOp}. The proof from \cite{HOp} can be easily
adapted for our case if we use Proposition 1 instead of the
Chevalley theorem about the invariants of the Weyl group.

In the case when all the multiplicities are integer we can
actually claim that all the solutions of the system \mref{sys} are
analytic {\it everywhere} but for this we will need some results
from \cite{CFV3} which we discuss in the next section.

\section{Baker-Akhiezer function and quantum integrals}\lb{sec62}

Let us assume now that all the parameters of the systems are
integer: for the system $\anm$ this means that $m\in \N$, and for
the system $\cnm$ all $m,l$ and $\kappa=\frac{2m+1}{2l+1}$ must be
positive integers.

According to \cite{CFV3} in that case there exists so-called {\it
Baker--Akhiezer function} related to such a configuration. This
function $\psi^\cA (k,x)$ has the form \beq{psideform}
\psi^\cA(k,x) = \left(P^\cA_N+\ldots +P^\cA_0\right) e^{(k,x)},
\eeq where $P^\cA_i=P^\cA_i(k,x)$ are some polynomials in $k, x$
of degree $i$ in $k$ and in $x$ with the highest term of the form
$$ P^\cA_N(k,x) = \prod_{\a\in \cA} (\a,k)^{m_\a} (\a,x)^{m_\a}.$$

The following {\it quasi-invariance condition} determines the
function $\psi^\cA$ uniquely (see \cite{CFV3}):
$$\psi^\cA(s_\a(k),x)-\psi^\cA(k,x)=o((\a,k)^{2 m_\a}) \
\mbox{near} \ (\a,k)=0, \,\, \a\in \cA, $$ where $s_\a$ is the
reflection with respect to the hyperplane $(\a,k)=0$.

Such a function $\psi^\cA$ exists only for very special class of
configurations including the Coxeter systems with invariant
integer multiplicities and the configurations $\anm$, $\cnm$ (see
\cite{ChV} for the latest results in this direction).

For any such configuration $\cA \subset V$ one can introduce the
following important notion. We will call a polynomial $q$ on $V$
{\it quasi-invariant for the system} $\cA$ if for all $\a\in \cA$
it is invariant up to order $2 m_\a$ with respect to the
reflections $s_\a$: \beq{quasidef} q(s_\a(k))=q(k) +
o\left((\a,k)^{2m_\a}\right) \eeq near the hyperplane $(\a,k)=0$
for any $\a\in \cA$. Equivalently we can say that for each $\a\in
\cA$ the odd normal derivatives $\p_\a^s q = (\a,\frac{\p}{\p
k})^s q$ vanish on the hyperplane $(\a,k)=0$ for $s=1,3,5,\ldots,
2m_\a-1.$

We denote the corresponding ring of quasi-invariants by $Q^\cA$.
We should mention that the terminology is slightly abusing. The
group generated by the reflections $s_\a$ is not a finite group
any more, and generally there are no polynomials except those of
$q(k)=k^2$ which are invariant with respect to all the reflections
$s_\a$. Nevertheless we will see that the rings $Q^\cA$ have some
nice properties similar to the Coxeter case.

The following general result explains the relation of this ring to
the generalized quantum CM problems. Let $D_{\cA}$ denote the ring
of differential operators which are regular outside the
hyperplanes  $(\a,x)=0, \a \in \cA.$

\begin{theorem}{\rm\cite{CV}}\lb{teorvsc}
If a system $\cA$ admits the Baker--Akhiezer function
$\psi^\cA(k,x)$, then there is a homomorphism $\chi^\cA: Q^\cA \to
D_{\cA}$ mapping a quasi-invariant $q(k)$ to the differential
operator $\cl_q(x,\frac{\p}{\p x})$ such that $$ { \cl}_q \psi^\cA
(k,x)=q(k)\psi^\cA (k,x). $$ Under this homomorphism the
quasi-invariant $k^2$ is mapped to the (gauged) generalized
Calogero--Moser operator $$ {\cal L} = \Delta -\sum_{\a\in
\cA}\frac{2m_\a}{(\a,x)}\p_\a. $$
\end{theorem}

In fact we can actually claim that the ring of quasi-invariants
$Q^\cA$ for our configurations $\anm$ and $\cnm$ is isomorphic to
the ring of {\it all} quantum integrals for the operator $\cl_\cA$
in the following sense. Denote by $D^\cA$ the maximal commutative
ring of differential operators with rational coefficients which
contains the operators $$ \cl_i=\chi^\anm(p_i) \quad \mbox{ for }
\cA=\anm, $$ and $$ \cl_i=\chi^\cnm(q_i) \quad \text{ for }
\cA=\cnm, $$
 $i=1,\ldots,n+1$. The following result can be proved similarly to
 the Coxeter case \cite{FV2}.

\begin{theorem}
The map $\chi^\cA$ is an isomorphism between the ring $Q^\cA$ and
the ring of quantum integrals $D^\cA$ of the corresponding
deformed CM systems (\ref{AA}), (\ref{CC}).
\end{theorem}

Another relation which will be important for us is the invariance
of the quasi-invariants under the action of all the quantum
integrals (cf. \cite{FV2}).

\begin{prop}
For the systems $\cA=\anm, \cnm$
the space $Q^\cA$ of quasi-invariants is invariant under the
action of the operators $\cl_q$, $q\in Q^\cA$.
\end{prop}
{\bf Proof.} We will use the result from \cite{CFV3} which says
that the operator $L_\cA$ preserves the space $\Phi$ of
meromorphic functions $\phi(x)$ such that $$ \phi(x)\prod_{\a\in
\cA} (\a,x)^{m_\a}  \quad \text{is holomorphic in} \quad \C^{n+1},
$$ and $$ \p_\a^{2s-1} (\phi(x) (\a,x)^{m_\a})|_{(\a,x)=0}=0 $$
for $s=1,\ldots,m_\a$, and  $\a\in \cA$ (see Lemma in the proof of
Theorem 3.1 in \cite{CFV3}). Since $\cl_\cA=\cl_{k^2}=g_\cA L_\cA
g_\cA^{-1}$ for $g=\prod_{\a\in A} (\a,x)^{m_\a}$, the operator
$\cl_\cA$ preserves the space $\widetilde\Phi$ consisting of the
functions $f(x)$ which are holomorphic in $\C^{n+1}$ and satisfy
the relations \beq{gm} \p_\a^{2s-1}
(f(x)\prod_{\b\ne\a}(\b,x)^{-m_\b})|_{(\a,x)=0}=0, \eeq
$s=1,\ldots,m_\a$, $\a\in \cA$.

Now we use the identities \mref{firstidlist} to claim that
\beq{XY} \p_\a^{2s-1}
(\prod_{\b\ne\a}(\b,x)^{-m_\b})|_{(\a,x)=0}=0, \eeq
$s=1,\ldots,m_\a$, $\a\in A$. Indeed, when $s=m_a=1$ these
relations are equivalent to \mref{firstidlist}. If $\a$ is such
that $m_a>1$ then $\prod_{\b\ne\a}(\b,x)^{-m_\b}$ is symmetric
with respect to reflection $s_\a$ and thus \mref{XY} obviously
holds.

Due to \mref{XY} the relations \mref{gm}
 can be rewritten as
 $$
 \p_\a^{2s-1} f(x)|_{(\a,x)=0} =0
 $$
 which means that $f(x)$ is quasi-invariant. Thus $\cl_\cA$
 preserves the space of quasi-invariant functions. Due to its
 form, $\cl_\cA$  also preserves the subspace of quasi-invariant
 polynomials. Now the Proposition follows from Berest's formula \cite{B}:
 $\cl_q=\mbox{const }
 ad_{\cl_\cA}^{\deg q} q$.

In particular we have
 \begin{cor}\label{edinitsa}
 For any $q\in Q^\cA$ one has $\cl_q 1 =0$.
 \end{cor}

\section{Structure of the rings of quasi-invariants}

In this section we are going to investigate the rings $Q^\cA$ of
quasi-invariants related to the configurations $\anm$, $\cnm$ in
more detail.

First of all we would like to mention that all the deformed Newton
sums $p_s$ and $q_s$ given by (\ref{ps}), (\ref{qs}) are the
quasi-invariants for the systems $\anm$ and $\cnm$ respectively.
Since they are symmetric with respect to the first $n$ coordinates
the quasi-invariance must be checked only for the hyperplanes with
multiplicity one, which is an easy calculation.

In particular the corresponding algebras of deformed invariants
$P^\cA$ generated by the first $n+1$ deformed Newton sums are the
subalgebras in $Q^\cA.$

\begin{theorem}
The algebras of quasi-invariants $Q^\cA$ are finitely generated.
\end{theorem}

Indeed by Proposition 1 the algebra of all polynomials is a
finitely generated free module over $P^\cA$ since positive
integers never belong to the exceptional set. To conclude the
theorem one can use a standard result from commutative algebra
(see e.g. Proposition 7.8 from Atiyah-Macdonald \cite{AM}).

We are going to prove that the algebras $Q^\cA$ are freely
generated as the modules over $P^\cA.$ This means that the rings
of quasi-invariants $Q^\cA$ have the Cohen-Macaulay property. We
will follow the scheme from Etingof-Ginzburg paper \cite{EG} where
a similar result for the Coxeter configurations was proved. A very
essential fact for the considerations in \cite{EG} is the lemma
claiming that the value of the Baker--Akhiezer function at zero is
nonzero. The proof was based on Opdam's results from \cite{O2} and
valid only in the Coxeter case.

We present now a different proof of this statement which works
both in Coxeter and in deformed cases.

\begin{prop}\lb{psine0}
The values of the Baker--Akhiezer functions $\psi^\cA$ related to
the systems $\cA=\anm$, $\cnm$ at zero are not zero: $$
\psi^\cA(0,0) \ne 0. $$
\end{prop}
{\bf Proof.} Consider the system of differential equations
\mref{sys}: \beq{sysproof} \cl_i \vf = \lambda_i \vf, \quad
i=1,\ldots,n+1, \eeq where $\cl_i=\chi^\anm(p_i)$ or
$\chi^\cnm(q_i)$. By Proposition \ref{prophol} we can fix a
solution $\vf=\vf(\lambda,x;x_0,a)$ locally at generic point $x_0$
by requirements $\g_i(\p) \vf(x_0)=a_i$, $i=1,\ldots,\mu$ and
consider its analytic continuation in $\C^{n+1}\times \C^{n+1}$.
The corresponding function $\vf$ is analytic at $x\notin \Sigma:
\{\cup_{\a\in\cA} (\a,x)=0\}$ and at arbitrary
$\lambda\in\C^{n+1}$.

On the other hand for generic $\lambda\in\C^{n+1}$ there exist
$\mu = \mu^\cA$ linearly independent solutions
$\psi_i(\lambda,x)=\psi^\cA(k^{(i)},x)$ to the system
\mref{sysproof} corresponding to different solutions $k^{(1)},
\ldots, k^{(\mu)}$ of the systems of the algebraic equations $
p_i(k)=\lambda_i, \quad i=1,\ldots,n+1$ (and respectively $
q_i(k)=\lambda_i, \quad i=1,\ldots,n+1$). Since by Proposition
\ref{prophol} the solution space to \mref{sysproof} has dimension
$\mu$, for generic $\lambda$ we have representation
\beq{repforsol} \vf(\lambda,x;x_0,a) = \sum_{i=1}^\mu c_i(\lambda)
\psi_i(\lambda,x), \eeq and in particular $\vf(\lambda,x;x_0,a)$
is holomorphic everywhere in $x$. By Hartogs theorem it follows
that $\vf(\lambda,x;x_0,a)$ is holomorphic in $\C^{n+1}\times
\C^{n+1}$.

Let us consider the value $\vf(\lambda,0;x_0,a)$. Suppose that $$
\psi^\cA(0,0)=\psi^\cA(k,0) = \psi^\cA(0,x) =0. $$ Then from
representation \mref{repforsol} we conclude that
$\vf(\lambda,0;x_0,a)=0$ for generic and therefore for all
$\lambda$. Now we choose $a=(1,0,\ldots,0)$. By Proposition
\ref{prophol} the solution $\vf=\vf(0,x;x_0,a)$ is uniquely
defined by the properties $\vf(x_0)=1, \g_i(\p) \vf(x_0)=0,
i=2,\ldots, \mu$. By Corollary \ref{edinitsa} we have that
$\vf(0,x;x_0,a)\equiv 1$, which is a contradiction with
$\vf(0,0;x_0,a)=0.$ This proves the Proposition.

Notice that as a by-product we have proved the following
\begin{prop}\lb{holo}
If all the multiplicities are integer all locally analytic
solutions of the system \mref{sys} are analytic everywhere in
${\bf C}^{n+1}$.
\end{prop}

Now everything is prepared for the next step, which is to define
the bilinear form on $Q^\cA$. For any $p,q \in Q^\cA$ we define
\beq{formadef} (p,q)^\cA = \cl_p q |_{x=0}, \eeq where the
operator $\cl_p=\chi^\cA(p)$ is the operator defined in Theorem
\ref{teorvsc}. As in the Coxeter case \cite{FV2}, \cite{EG} this
form defines a scalar product on $Q^\cA$.

Let $c_\cA$ be the value of the Baker-Akhiezer function at zero:
$\psi^\cA(0,0)= c_\cA \ne 0$ by Proposition \ref{psine0}.

\begin{theorem}
For the systems $\cA=\anm, \cnm$ bilinear form  \mref{formadef} is
symmetric and non-degenerate. It can be written as
\beq{anotherformdef} (p,q)^\cA = c^{-1}_\cA \cl_p^{(x)}\cl_q^{(k)}
\psi^\cA(k,x) |_{k=x=0}. \eeq With respect to this form the
differential operator $\cl_q$ is adjoint to the operator of
multiplication by $q\in Q^\cA.$
\end{theorem}

\noindent {\bf Proof.} We closely follow \cite{EG} here. Denote by
$g_i(k)$ any homogeneous basis in the space $Q^\cA$ of
quasi-invariants. Consider the Baker--Akhiezer function
$\psi^\cA(k,x)$. Since it satisfies the quasi-invariant
conditions, we can write its Taylor series in the following form
\beq{psirazl} \psi^\cA(k,x) = \sum_{i=0}^\infty g_i(k) g^i(x),
\eeq where $g^i(x)$ are some polynomials. Indeed, we may think of
\mref{psirazl} as of decomposition of $\psi^\cA$ via the basis
$g_i(k)$. Then coefficients $g^i(x)$ of this decomposition turn
out to be the quasi-invariant polynomials as well. Indeed, as it
is established in \cite{CFV3} the function $\psi^\cA(k,x)$ is
symmetric: \beq{psisym} \psi^\cA(k,x)=\psi^\cA(x,k). \eeq Hence it
satisfies the quasi-invariant conditions in $x$ variables as well.
Since $g_i(k)$ is a basis, each polynomial $g^i(x)$ must satisfy
quasi-invariant conditions. Now for a given quasi-invariant $q(k)$
consider  the following equations $$ q(k)\psi^\cA(k,x)=\cl_q
\psi^\cA(k,x) = \sum_i g_i(k) \cl_q g^i(x). $$ We put $x=0$ and
use the fact that $\psi^\cA(k,0) = \psi^\cA(0,0)= c_\cA$ to obtain
the relation $$ c_\cA q(k) = \sum_i g_i(k) \cl_q g^i|_{x=0}. $$
Since $g_i(k)$ is a basis in $Q^\cA$, taking $q(k)=g_j(k)$ we get
$$ \cl_{g_j} g^i|_{x=0} = c_\cA \delta^i_j. $$ This proves that
the bilinear form \mref{formadef} is non-degenerate.

Now the formula \mref{anotherformdef} simply follows: for any
$p,q\in Q^\cA$ we have $$ \cl_p^{(x)}\cl_q^{(k)} \psi^\cA(k,x)
|_{k=x=0} = \cl_p^{(x)} \left(q(x) \psi^\cA(k,x)\right)|_{k=x=0} =
 c_\cA \cl_p^{(x)} \left(q(x)\right)|_{x=0} =  c_\cA (p,q)^\cA. $$ The
symmetry of the bilinear form follows from the symmetry
\mref{psisym} of $\psi^\cA$: $$ (q,p)^\cA = c_\cA^{-1}
\cl_q^{(x)}\cl_p^{(k)} \psi^\cA(k,x) |_{k=x=0} = c_\cA^{-1}
\cl_p^{(k)}\cl_q^{(x)} \psi^\cA(k,x) |_{k=x=0} = $$ $$ =c_\cA^{-1}
\cl_p^{(k)}\cl_q^{(x)} \psi^\cA(x,k) |_{k=x=0} = (p,q)^\cA. $$

For arbitrary three quasi-invariants $p, q, r \in Q^\cA$ we have
$$ ( q r, p)^\cA = \cl_{q r} p |_{x=0} = \cl_r \cl_q p|_{x=0} =
(r, \cl_q p)^\cA, $$ so the operator of multiplication by $q$ is
adjoint to the operator $\cl_q$. This completes the proof of the
theorem.

As in the Coxeter case \cite{EG} the defined form can be extended
to the pairing between $Q^\cA$ and the formal series of
quasi-invariants $\hat Q^\cA$ still being non-degenerate.

Consider now the space $H^\cA(\lambda)$ which consists of formal
series solutions to the corresponding system \mref{sys}.

\begin{lemma}
For any $\lambda\in \C^{n+1}$ the following inclusions hold
$$
H^\anm(\lambda) \subset \hat Q^\anm, \quad H^\cnm(\lambda) \subset \hat
Q^\cnm.
$$
\end{lemma}

\noindent {\bf Proof} We actually will prove a stronger statement.
Let some power series $f(x)$ satisfy the equation $$ \cl_\cA f = E
f, \quad E \in \C $$ where $\cl_\cA$ is the deformed
Calogero--Moser operator. We claim that $f(x)$ satisfies the
quasi-invariance conditions (c.f. \cite{CFV3}).

Let $L_\cA =g^{-1}\cl_\cA  g, \quad g=\prod_{\a\in \cA}
(\a,x)^{m_\a}$ be the potential form of the operator $\cl_\cA $,
then we have the equation \beq{lgf} L_\cA  \left(g^{-1}f\right) =
E g^{-1}f. \eeq Consider the expansions at $(\a,x)=0$, $\a\in
\cA$, where $\a$ is normalized to have unit length:
\begin{gather}
g^{-1}f = (\a,x)^{-m_\a}\left(f_0+f_1 (\a,x) + f_2 (\a,x)^2+\ldots
\right),\label{gf1}\\[3mm]\label{gf2} L_\cA =\p^2_\a+\tilde\Delta
-\frac{m_\a(m_\a+1)}{(\a,x)^2} - u_0 - u_1(\a,x) - u_2(\a,x)^2 -
\ldots,
\end{gather}
where $\tilde\Delta$ is the Laplacian in the hyperplane
$(\a,x)=0$, and all the coefficients $f_i, u_i$ are functions on
the hyperplane. One can check by direct calculation that for the
systems $\anm$, $\cnm$ $$ u_1=u_3=\ldots=u_{2m_\a-1}=0. $$
Substitution of the expansions \mref{gf1}, \mref{gf2} into
equation \mref{lgf} shows that $$ f_1=f_3=\ldots=f_{2m_\a-1}=0, $$
so that \beq{tempf} f =\prod_{\b\ne\a}(\b,x)^{m_\b} \left(f_0+ f_2
(\a,x)^2+\ldots \right). \eeq We have to show that \beq{deriv}
\p_\a^s f =0 \quad \mbox{ at } \, (\a,x)=0 \eeq for $s=1,3,\ldots,
2m_\a-1$. Since the systems $\anm$, $\cnm$ are symmetric with
respect to each hyperplane with multiplicity $m_\a>1$ for such
hyperplane these conditions are obviously satisfied. If the
multiplicity $m_\a=1$ then the condition \mref{deriv} follows from
the form \mref{tempf} and the relation $$ \p_\a
\left(\prod_{\b\ne\a}(\b,x)^{m_\b}\right) =0  \quad \mbox{ if } \,
(\a,x)=0, $$ which is equivalent to identity \mref{firstidlist}.
Lemma is proven.

Let us introduce now the following ideal $I^\cA(\lambda)\subset
Q^\cA$ generated by the polynomials $p_s-\lambda_s, \quad
s=1,\ldots,n+1$ in the case $\anm$ and by the polynomials
$q_s-\lambda_s, \quad s=1,\ldots,n+1$ in $\cnm$ case.

\begin{prop}\lb{propdimdef}
The following two dimensions are equal
\beq{eqdimdef}
\dim Q^\cA/I^\cA(\lambda) = \dim H^\cA(\lambda),
\eeq
where $\cA=\anm, \cnm.$
\end{prop}
The proof follows from the following two lemmas.

\begin{lemma}
The orthogonal complement to the ideal
$I^\cA(\lambda)$ in the
completion $\hat Q^\cA$ is the solution space
$H^\cA(\lambda)$.
\end{lemma}

\begin{lemma}
The space $H^\cA(\lambda)$ is isomorphic to the dual space
$(Q^\cA/I^\cA(\lambda))^{*}$. The element $h\in H^\cA(\lambda)$
defines a functional
on the factor space by the formula
$$
q+I^\cA(\lambda) \to (q+I^\cA(\lambda),h)
$$
where $q+I^\cA(\lambda)$ is arbitrary element from $Q^\cA/I^\cA(\lambda)$.
\end{lemma}
The proof is based on the non-degeneracy of the scalar product
\mref{formadef}.

The next claim is that the dimension of $Q^\cA/I^\cA(\lambda)$ is
the largest when $\lambda=0$.
\begin{lemma}\lb{lemmochka6def}
For any $\lambda\in\C^{n+1}$ the inequality
$$
\mathrm{dim\,} Q^\cA/I^\cA(0) \ge \mathrm{dim \,} Q^\cA/I^\cA(\lambda)
$$
holds. If $r_1, \ldots, r_{N}$ is a homogeneous basis in the
complement  to the ideal $I^\cA(0)$ then the classes $\bar r_i = r_i
+I^\cA(\lambda)$ generate $Q^\cA/I^\cA(\lambda)$.
\end{lemma}

\noindent {\bf Proof} Consider a homogeneous basis $r_1, \ldots,
r_{N}$ in the complement to the ideal $I^\cA(0)$. For arbitrary
$q\in Q^\cA$ we have $$ q = \sum_{i=1}^N  \mu_i r_i + q', $$ where
$q' \in I^\cA(0)$ and $\mu_i$ are some constants. Since $q' \in
I^\cA(0)$, for $q'$ there is a representation $$ q' =
\sum_{i=1}^{n+1} t_i p_i $$ when $\cA=\anm$, and one should write
polynomials $q_i$ instead of $p_i$ for the system $\cA=\cnm$. The
polynomials $t_i\in Q^\cA$ have the degrees less than $\deg q$,
and
\beq{4110} q=\sum_{i=1}^N  \mu_i r_i+\sum_{i=1}^{n+1}t_i p_i. \eeq
For $t_i$ we have a similar representation \beq{4111}
t_i=\sum_{j=1}^N \mu_{ij} r_j + f_i \eeq with $f_i \in I^\cA(0)$.
Next we should write $f_i$ as a combination of $p_1, \ldots,
p_{n+1}$ and substitute \mref{4111} back to \mref{4110}.
Continuing the process we arrive to the presentation \beq{rel}
q=\sum_{i=1}^N \tau_i r_i, \eeq where $\tau_i$ are polynomials in
$p_1, \ldots, p_{n+1}$. Now, consider relation \mref{rel} modulo
ideal $I^\cA(\lambda)=(p_i-\lambda_i)$. We get $$ \bar q = \sum
c_i \bar r_i, $$ where the constants $c_i=\tau_i(\lambda)$. Thus
we see that $\bar r_i$ generate the whole space
$Q^\cA/I^\cA(\lambda)$, and hence $\dim Q^\cA/I^\cA(\lambda)\le
N$. The lemma is proven.

\begin{prop}\lb{razmravny}
For any $\lambda\in\C^{n+1}$ the dimensions
$$
\dim Q^\cA/I^\cA(\lambda)=\dim Q^\cA/I^\cA(0) <\infty.
$$
\end{prop}
{\bf Proof.} We will use Proposition 6. Recall that the spaces
$H^\cA(\lambda)$ consist of formal solutions to the system
\mref{sys}. But by Proposition \ref{holo} any locally analytic
solution of \mref{sys} is actually holomorphic everywhere in
$\C^{n+1}$. This implies that $$ \dim H^\cA(\lambda)\ge \mu^\cA,
$$ where $\mu^\cA$ denotes the dimension of the space of (locally)
analytic solutions.

Consider now the value $\lambda=0$.  We claim that all the formal
series solutions $F$ to systems \mref{sys} with $\lambda=0$ are
actually polynomial. Indeed, if $F\in\C[[x]]$ is a formal solution
then every homogeneous component of $F$ is also a solution to the
system due to homogeneity of system \mref{sys} if $\lambda=0$. So
if a formal solution exists then there are infinitely many
analytic solutions. But the space of analytic solutions has finite
dimension $\mu^\cA$. Therefore $\dim H^\cA(0)$ is also equal to
$\mu^\cA$.

Thus using Lemma \ref{lemmochka6def} and Proposition
\ref{propdimdef} we have $$ \mu^\cA = \dim Q^\cA/I^\cA(0) \ge \dim
Q^\cA/I^\cA(\lambda)\ge \mu^\cA, $$ which proves the statement.

Finally we have the main result of this section.
\begin{theorem}\lb{CMdefo}
The algebra of quasi-invariants $Q^\cA$ is a free module of rank
$\mu^\cA$ over its polynomial subalgebra $P^\cA.$
\end{theorem}
{\bf Proof} Consider a homogeneous basis $r_1, \ldots, r_{\mu}$ in
the complement to the ideal $I^\cA(0)$. As we established in the proof of Lemma
\ref{lemmochka6def}, every element $q\in Q^\cA$ has a
representation $$ q=\sum_{i=1}^\mu \tau_i r_i, $$ where  $\tau_i$
are some polynomials in $p_1, \ldots, p_{n+1}$, and the classes
$\bar r_i = r_i +I^\cA(\lambda)$ generate the space
$Q^\cA/I^\cA(\lambda)$.
 From Proposition \ref{razmravny} it follows that the elements
$\bar r_i$ are linearly independent. Now, if the polynomials $r_1,
\ldots, r_\mu$ were dependent over the ring generated by
$p_1,\ldots,p_{n+1}$ we would have a relation $$ \sum_{i=1}^\mu
s_i r_i =0 $$ where $s_i$ are some polynomials in
$p_1,\ldots,p_{n+1}$. Hence, taking this relation modulo
$I^\cA(\lambda)$ we get $$ \sum_{i=1}^\mu c_i \bar r_i =0 $$ where
$c_i=s_i(\lambda)$. Since generically all $c_i$ are nonzero, this
contradicts to linear independence of $\bar r_i$. The theorem is
proven.

\section{Poincare series for quasi-invariants}\lb{sec63}

In this section we calculate the Poincare series for
two-dimensional deformations $\adva$ and $\cdva$.

Let $P^\cA(t)$ be the Poincare polynomial for the complement to
the ideal $I^\cA(0)$ in $Q^\cA$. If $p^\cA(t)$ is the Poincare
series for the quasi-invariants then from Theorem \ref{CMdefo} it
follows that \beq{poina2}
p^\adva(t)=\frac{P^\adva(t)}{(1-t)(1-t^2)(1-t^3)}, \eeq
\beq{poinc2} p^\cdva(t)=\frac{P^\cdva(t)}{(1-t^2)(1-t^4)}. \eeq We
are going to compute the polynomials $P^\adva(t)$ and
$P^\cdva(t).$

\begin{theorem}\lb{poincarec2}
The Poincare polynomials for $\adva$ and $\cdva$ have the form:
\beq{pa2formula} P^\adva(t)=1+t^4+t^5+t^{2m+2}+t^{2m+3}+t^{2m+7},
\eeq \beq{pc2formula}
P^\cdva(t)=1+t^6+t^{2m+3}+t^{2m+5}+t^{2l+3}+t^{2l+5}+t^{2(m+l+1)}+t^{2(m+l+4)}.
\eeq
\end{theorem}
Since these Poincare polynomials are palindromic according to the
general Stanley result \cite{Stanley} we have the following result
which we believe to be true in arbitrary dimension $n.$

\begin{cor}\label{Gorenstein}
 The rings $Q^\adva$ and $Q^\cdva$ are Gorenstein.
\end{cor}
{\bf Proof of the Theorem.} We first consider $\cdva$ case. We
will actually compute the Poincare series for $Q^\cdva$ by direct
computation.

Consider an arbitrary polynomial of degree $n$ in two variables :
$$ q(x,y)= \sum_{i=0}^n a_i x^i y^{n-i}. $$ The quasi-invariance
conditions on the lines $x=0$, $y=0$ have the following simple
form: \beq{quasix} a_1=a_3=\ldots=a_{2m-1}=0, \eeq and
\beq{quasiy} a_{n-1}=a_{n-3}=\ldots=a_{n-(2l-1)}=0. \eeq We have
two more quasi-invariance conditions on the lines $\Pi_\pm: \xi x
\pm y=0$, where \beq{ksi} \xi=\sqrt{\frac{2l+1}{2m+1}}. \eeq Let
us write these conditions
\begin{multline*}
(\xi \p_x \pm  \p_y) q(x,y)|_{\Pi_\pm} = \sum_{i=0}^n (\xi i a_i
x^{i-1}y^{n-i}\pm  (n-i)a_i x^i y^{n-i-1})|_{\Pi_\pm}=\\
\sum_{i=0}^n
(\xi i a_i (\mp\xi)^{n-i}\pm (n-i)a_i
(\mp\xi)^{n-i-1})x^{n-1}|_{\Pi_\pm}=\\
\sum_{i=0}^n
(\mp i a_i (\mp\xi)^{n-i+1}\pm (n-i)a_i
(\mp\xi)^{n-i-1})x^{n-1}|_{\Pi_\pm}=
0.
\end{multline*}
Thus we obtain
$$
\sum_{i=0}^n
a_i (\pm\xi)^{n-i-1}\left((n-i)-\xi^2 i\right)=0.
$$
These equations are equivalent to the following
\begin{gather}
\lb{quasilines1}
\sum_{\nad{i=0}{i=2k}}^n
a_i \xi^{n-i-1}\left((n-i)-\xi^2 i\right)=0, \\ \lb{quasilines2}
\sum_{\nad{i=0}{i=2k+1}}^n
a_i \xi^{n-i-1}\left((n-i)-\xi^2 i\right)=0.
\end{gather}
Now the question is: how many linear independent conditions on the
coefficients $a_0,\ldots,a_n$ are among \mref{quasix},
\mref{quasiy}, \mref{quasilines1}, \mref{quasilines2} ?

Consider first the case when $n$ is odd. The equations on the
coefficients $a_i$ split into the equations for the coefficients
with odd and even indices. Due to the equations \mref{quasix},
\mref{quasiy} we have $\frac{n+1}2-l$ nontrivial even
coefficients, and $\frac{n+1}2-m$ nontrivial odd coefficients.
Both equations \mref{quasilines1}, \mref{quasilines2} are
nontrivial, each of them is a linear equation for the nontrivial
coefficients $a_i$ with even, and, correspondingly, odd, indices
$i$. Summarizing,  for a given odd $n$ we have $\frac{n+1}2-l-1$ -
dimensional space of the coefficients with even indices, and
$\frac{n+1}2-m-1$ - dimensional space of the coefficients with odd
indices. Therefore the Poincare series for the quasi-invariants
with odd degrees has the form
\begin{multline}\lb{podd}
p_{odd}=\sum_{\nad{n\ge 2(l+1)+1}{n=2k}}
\left(\frac{n+1}2-l-1\right)t^n +
\sum_{\nad{n\ge 2(m+1)+1}{n=2k}}
\left(\frac{n+1}2-m-1\right)t^n=\\[5mm]
t^{2l+3}(1+2t^2+3t^4+4t^6+\ldots)+
t^{2m+3}(1+2t^2+3t^4+4t^6+\ldots)=\\[5mm]
\frac{t^{2l+3}+t^{2m+3}}{(1-t^2)^2}.
\end{multline}
Now let us consider the even part of the Poincare series. When $n$
is even conditions \mref{quasix}, \mref{quasiy} are both the
conditions for the coefficients $a_i$ with odd indices $i$.
Therefore for the coefficients $a_i$ with even $i$ the only
restriction is given by \mref{quasilines1}, and it is nontrivial
unless $n=0$. Thus the space of possible even coefficients has the
dimension $\frac{n}2+1-1=\frac{n}2$ if $n\ne 0$. If $n=0$ we still
have one-dimensional space of quasi-invariants given by the
constants $a_0$. The conditions on the odd coefficients $a_i$ are
independent from the conditions on the even coefficients.
Therefore the Poincare series $p_{even}$ of even degree
quasi-invariants has the form $$
p_{even}=p_{even}^{odd}+p_{even}^{even}. $$ As we just analyzed,
\beq{pevev} p_{even}^{even}=1+\sum_{\nad{n=2k}{n>0}} \frac{n}2 t^n
= 1+t^2(1+2 t^2 + 3t^4+\ldots) = 1+\frac{t^2}{(1-t^2)^2}. \eeq We
are left to calculate the Poincare subseries $p_{even}^{odd}$
corresponding to quasi-invariants which have even degree and which
are odd in $x$ and $y$. We have $\frac{n}2$ odd coefficients $a_1,
a_3, \ldots, a_{n-1}$ and  conditions \mref{quasix},
\mref{quasiy}, \mref{quasilines2}. If $m+l\ge \frac{n}2$ then by
conditions \mref{quasix}, \mref{quasiy} all the coefficients must
be zero and there are no quasi-invariants of the form required. If
$m+l=\frac{n}2-1$ then we have $m+l$ vanishing conditions
\mref{quasix}, \mref{quasiy}, and equation \mref{quasilines2} for
the only nontrivial coefficient $a_{2m+1}$. But $$ (n-i)-\xi^2 i
=0 $$ if $i=2m+1$ and $m+l=\frac{n}2-1$ (see the form \mref{ksi}).
Hence equation \mref{quasilines2} gives no additional
restrictions, and there exists one dimensional space of
quasi-invariants when $n$ satisfies $m+l=\frac{n}2-1$, i.e. when
$n=2(m+l+1)$. If $n>2(m+l+1)$ then the equation \mref{quasilines2}
is already nontrivial and we have the dimension of
quasi-invariants being equal to $\frac{n}2-m-l-1$. Thus we have
\begin{multline}\lb{pevod}
p_{even}^{odd}=t^{2(m+l+1)}+\sum_{k=0}^\infty k t^{2(m+l+1)+2k}=\\
t^{2(m+l+1)}+t^{2(m+l+1)+2}\left(1+2t^2+3t^4+\ldots\right)=
t^{2(m+l+1)}+\frac{t^{2(m+l+1)+2}}{(1-t^2)^2}
\end{multline}
Altogether collecting \mref{podd}, \mref{pevev} and \mref{pevod}
we get the following expression for the Poincare series
\begin{multline*}
p=\frac{t^{2l+3}+t^{2m+3}}{(1-t^2)^2}+ 1+\frac{t^2}{(1-t^2)^2} +
t^{2(m+l+1)}+\frac{t^{2(m+l+1)+2}}{(1-t^2)^2}=\\[5mm]
=\frac{1+t^6+t^{2l+3}+t^{2m+3}+t^{2l+5}+t^{2m+5}+ t^{2(m+l+1)} +
t^{2(m+l+4)}}{(1-t^2)(1-t^4)}.
\end{multline*}
This proves the formula \mref{pc2formula}.

The case $\adva$ can be reduced to the previous one. The
observation is that the system $\adva$ considered as a system in
$\C^2\simeq \{x\in\C^3| x_1+x_2+\frac1{\sqrt{m}} x_3=0\}$ is
identical to $\cdva$ with multiplicity $l=0$. Since the Poincare
series for the quasi-invariants in $\C^2$ and $\C^3$ satisfy
obvious relation $$ \frac{p^{\adva\subset\C^2}}{1-t} =
p^{\adva\subset\C^3} $$ we have $$
p^{\adva\subset\C^2}=(1-t)p^{\adva\subset\C^3}=\frac{P^{\adva}}{(1-t^2)(1-t^3)}
=
\frac{P^{{\cal C}_2(m,0)}}{(1-t^2)(1-t^4)}. $$ Thus we have $$
P^{{\cal C}_2(m,0)} (1-t^3)= P^{\adva} (1-t^4), $$ which together
with \mref{pc2formula} implies the formula \mref{pa2formula}.

{\bf Remark.} Note that the "beginnings" of the Poincare
polynomials \mref{pa2formula}, \mref{pc2formula} do not depend on
the multiplicities. We conjecture that the corresponding "stable"
parts of Poincare polynomials for the general $n$ have the form
$$P_{stable}^{\anm} = 1 + t^{n+2} + t^{n+3} + \dots + t^{2n+1}$$
and $$P_{stable}^{\cnm} = 1 + t^{2(n+2)} + t^{2(n+3)} + \dots +
t^{2(2n+1)}.$$ This form is motivated by the formula (24) from
\cite{SV} for the Poincare series for the subalgebra of
quasi-invariants generated by {\it all} deformed Newton sums for
generic values of deformation parameter. It is natural also to
suggest that for large multiplicities $m$ the corresponding $n$
deformed Newton sums $p_s$ with $s = n+2, \dots, 2n+1$ can be
chosen as the first $n$ free generators for the module $Q^\anm$
over $P^\anm.$

\section{Deformed $m$-harmonic polynomials}\lb{sec64}

In the paper \cite{FV2} we introduced the notion of $m$-harmonic
polynomials related to any Coxeter configuration. Their deformed
versions related to configurations $\anm$ and $\cnm$ are defined
in a similar way as the solutions of the system \beq{m-sys} \cl_i
f = 0, \quad i=1,\ldots, n+1, \eeq where $\cl_i$ are the operators
from the Theorem \ref{teor1}.

\begin{theorem}  All the solutions of the system (\ref{m-sys}) are
polynomial and moreover are quasi-invariants. The degrees of a
basis of homogeneous solutions to (\ref{m-sys}) coincide with the
degrees of homogeneous generators of $Q^\cA$ as a module over
$P^\cA$.
\end{theorem}
The proof follows from the results of section 4. Indeed, the first
part follows from Lemma 2, Propositions 5 and the homogeneity of
the system (\ref{m-sys}). To prove the second part note that
according to Lemma 4 the space of the deformed $m$-harmonic
polynomials $H^\cA(0)$ is dual to the quotient $Q^\cA/I^\cA(0)$
and use the homogeneity of the bilinear form \mref{formadef}.

In the Coxeter case we have conjectured in \cite{FV1} that these
polynomials can be chosen as free generators for the corresponding
algebra of quasi-invariants over invariants (like in the classical
case due to Chevalley and Steinberg \cite{Chev, St}). Although
this turned out to be false in general (Etingof and Ginzburg
\cite{EG} found a counterexample related to $B_6$ system) the
question how often this fails is not clear yet (see \cite{FV2} for
the discussion of this). It is interesting therefore to look what
happens in the deformed case. In this section we consider only the
simplest case $\adva$.

Let us consider the system $\adva$ as a system in $\C^2$
consisting of the vectors $e_1$ with multiplicity $m$ and
$e_1\pm\sqrt{2m+1}e_2$ with multiplicity 1. In this embedding the
first deformed Newton sums have the form $$ p_1=x^2+y^2, \quad p_2
= (2m-1)y^3-3x^2 y. $$ The corresponding quantum integrals are $$
\cl_{1} = \p^2_x+\p^2_y+\frac{4(1+2m)y}{x^2-(2m+1)y^2}\p_y+
\frac{2m(1+2m)y^2-2(2+m)x^2}{x(x^2-(2m+1)y^2)}\p_x, $$ and
\begin{multline}
\cl_{2} = (2m-1)\p_y^3-3\p^2_x\p_y
-\frac{6(4m^2-1)y}{(2m+1)y^2-x^2}\p^2_y
+\frac{6(2m+1)y}{(2m+1)y^2-x^2}\p^2_x\\[5mm]
+\frac{6((m+2)x^2-m(2m+1)y^2)}{x(x^2-(2m+1)y^2)}\p_x\p_y
\\[5mm]
-\frac{12(2m+1)y((m-2)x^2+m(1+2m)y^2)}{x(x^2-(2m+1)y^2)^2}\p_x
-\frac{12(2m+1)(x^2+(1-4m^2)y^2)}{(x^2-(2m+1)y^2)^2}\p_y.
\end{multline}
The deformed $m$-harmonic polynomials are solutions to the system
\begin{equation}
\begin{cases}
\cl_{1}f=0\\
\cl_{{2}}f=0.
\end{cases}
\end{equation}

\begin{prop}\lb{m-harm}
The space of the deformed $m$-harmonic polynomials for the
configuration $\adva$ is generated by the following 6
quasi-invariants:
\begin{multline*}
q_1=1, \quad q_2= x^4 +2(2m+1)x^2 y^2 + (1-4m^2)y^4, \quad q_3=
x^4 y - \frac15 (2m-3)(2m+1)y^5,\\[4mm] q_4=x^{2m+1}y, \quad
q_5=x^{2m+1}(x^2+(2m+3)y^2),\\[4mm] q_6= 5 x^{7 + 2 m} + x^{5 + 2
m} (-35 y^2 - 10 m y^2) +
     x^{3 + 2 m} (35 y^4 + 80 m y^4 + 20 m^2 y^4) +\\[4mm]
     x^{1 + 2 m} (-21 y^6 - 62 m y^6 - 44 m^2 y^6 -
           8 m^3 y^6).
\end{multline*}
\end{prop}

Analysis of these formulas shows that the deformed $m$-harmonics
fail to give a basis for the quasi-invariants module already for
the first non-trivial case $m=2.$ Indeed, in that case the
polynomial $q_3$ has the form $$q_3=y(x^2-y^2)(x^2+y^2) =-\frac13
p_1 p_2 $$ and thus belongs to the ideal generated by $p_1$ and
$p_2$.

Note that the degrees of $q_i$ are $$ 0,\, 4,\, 5,\, 2m+2,\, 2m+3,
\, 2m+7 $$ which is in agreement with Theorem \ref{poincarec2}.
When $m=1$ (i.e. in the Coxeter case) these degrees are known for
all $n$ (see \cite{FeV}).

\section{Acknowledgements}
We are grateful to Yu. Berest, O. Chalykh, P. Etingof and A.N.
Sergeev for useful discussions. The second author (A.P.V.) is
grateful to IHES (Bures-sur-Yvette, France) for the hospitality in
February 2003 when the final version of this paper was prepared.

This work was partially supported by EPSRC (grant GR/M69548).


\end{document}